\def\sun{$_\odot$}  
\documentstyle[12pt,epsf,epsfig]{article}     

\def\alwaysmath#1{{\ifmmode{#1}\else{$#1$}\fi}}
\def\he#1{\hbox{\alwaysmath{{}^{#1}}{\rm He}}}

\def\li#1{\hbox{\alwaysmath{{}^{#1}}{\rm Li}}}

\def\hii{H\thinspace{$\scriptstyle{\rm II}$}}
\def\etal{{\it et al.}~}
\def\beginapjbib{\begingroup \section*{\large \bf References}
   \parskip=.5ex plus 1.0pt
   \def\bibitem{\par \noindent \hangindent\parindent
      \hangafter=1}}
\def\endapjbib{\par \endgroup}
\def\endapjbib{\par \endgroup}
\def\beginfig{\begingroup \section*{\large \bf Figure Captions}
         \parskip=.5ex plus 1.0pt
         \def\figitim{\par \noindent \hangindent\parindent
                \hangafter=1}}
\def\endfig{\par \endgroup}
\begin{document}
\begin{titlepage}
\pagestyle{empty}
\baselineskip=21pt
\rightline{UMN-TH-1516/96}
\rightline{astro-ph/9610039}
\rightline{October 1996}
\vskip .2in
\begin{center}
{\large{\bf Low Mass Stars and the \he3 Problem}}
\end{center}
\vskip .1in
\begin{center}

Keith A. Olive$^1$,
David N. Schramm $^{2,3}$,
Sean T. Scully$^1$, 
and James W. Truran$^2$

$^1${\it School of Physics and Astronomy, University of Minnesota}
{\it Minneapolis, MN 55455}

$^2${\it Department of Astronomy and Astrophysics, Enrico Fermi Institute,
The University of Chicago, Chicago, IL  60637-1433}

$^3${\it NASA/Fermilab Astrophysics Center,
Fermi National Accelerator Laboratory, Batavia, IL  60510-0500}

\vskip .1in

\end{center}
\vskip .2in
\centerline{ {\bf Abstract} }
\baselineskip=18pt

The prediction of standard chemical evolution models of higher
abundances of \he3 at the solar and present-day epochs than are
observed indicates a possible problem with the yield of \he3 for stars
in the range of 1-3 $M_{\odot}$. Because \he3 is one of the nuclei 
produced in Big Bang Nucleosynthesis (BBN), it is noted that 
galactic and stellar evolution uncertainties necessarily
relax constraints based on \he3. We incorporate
into chemical evolution models which include outflow, 
the new yields for \he3 of Boothroyd \& Malaney (1995)
which predict that low mass stars are net destroyers of \he3.
Since these yields do not account for the high \he3/H ratio
observed in some planetary nebulae, we also consider the 
possibility that some fraction of stars in the 1 -- 3 M\sun\ 
range do not destroy their \he3 in theirpost main-sequence phase.  
We also consider the possibility that the gas expelled by stars 
in these mass ranges does not mix with the ISM 
instantaneously thus delaying the \he3 produced in these stars,
according to standard yields, from reaching the ISM. 
In general, we find that the Galactic D and \he3 
abundances can be fit regardless of whether
the primordial D/H value is high ($2 \times 10^{-4}$) or low 
($2.5 \times 10^{-5}$).

\baselineskip=18pt

\noindent
\end{titlepage}
\baselineskip=18pt                   
\def\la{~\mbox{\raisebox{-.7ex}{$\stackrel{<}{\sim}$}}~}
\def\ga{~\mbox{\raisebox{-.7ex}{$\stackrel{>}{\sim}$}}~}
\def\beq{\begin{equation}}
\def\eeq{\end{equation}}

\section{Introduction}                  

The abundances of the light elements, D, \he3, \he4, and \li7, serve
as a critical test to the standard hot big bang model (Walker \etal 1991).
In the case of two of these elements, \he4 and \li7,
primordial values can be reasonably well determined directly from observations.
\he4 may be inferred from low metallicity \hii\ regions 
(see e.g. Pagel \etal 1992; Olive \& Steigman 1995; Olive, Skillman, 
\& Steigman 1996; Skillman \etal
1996; Olive \& Scully 1996),
while the  uniform abundance of \li7 in halo dwarfs is interpreted to
be the primordial value for \li7/H (Spite \& Spite, 1982).  Although 
possible stellar depletion of as much as a factor of 2 can not be
categorically excluded, higher depletion is in conflict with \li6
observed in these stars and with maintaining the tightness of the
plateau (Steigman \etal 1993; Vauclair \& Charbonnel 1995; Lemoine \etal 1996).  
Fields \& Olive (1996) and Fields \etal (1996) have argued that on the
basis of these two isotopes alone, one can constrain the single parameter
(the baryon-to-photon ratio, $\eta$) of standard big bang
nucleosynthesis (SBBN) with the degree of constraint depending on the
allowed \li7 depletion and the maximum systematic errors allowed in the
\he4 determination.

The consistency of the two remaining elements, D and \he3, with such a
best fit $\eta$ based on \he4 and \li7, may pose a challenge for many 
chemical evolution models.
Let us first discuss deuterium which may eventually give the cleanest
constraints but at present has some ambiguities.
 Deuterium evolution by itself is straight
forward since D is only produced in BBN and destroyed by stellar 
processing (Reeves \etal 1973; Epstein \etal 1976).  Thus the present 
HST determined ISM D/H values give a robust upper bound on
$\eta$, but alone do not specify the degree of depletion between BBN and the
present day. While progress has been made recently in determining this
primordial value by observing D/H in quasar absorption systems, a 
consistent D$_p$ has not yet been found.
Both a high primordial value of D/H around $2 \times 10^{-4}$ (Carswell \etal
1994; Songaila \etal 1994; Rugers \& Hogan 1996a,b) 
and low D$_p$ of D/H $\sim 2.5 
\times 10^{-5}$ (Tytler, Fan \& Burles 1996; Burles \& Tytler 1996) have been
measured.  Both of these measurements have strengths and weaknesses.  The high D
measurement agrees well with best-fit $\eta$ for \he4 and \li7 (Fields \etal
1996) but requires destruction factors of $\sim 10$ for
D to reach its present-day observed values.  While models exist which
can accomplish this (see e.g. Vangioni-Flam \& Audouze 1988; 
Scully \& Olive 1995; Scully \etal 1996a,b), these models prove
problematical for \he3 since in traditional low mass star models 
D is converted into \he3.  The
low D measurement on the other hand proves less of a problem for 
traditional \he3 evolution
but would require large systematic errors on the primordial \he4
measurement (Copi, Schramm \& Turner 1995a) 
and a significantly higher primordial
\li7 abundance (by a factor of about 3). The possibility that both the high and
low D/H measurements are correct and indicate a possible baryon inhomogeneity
has been explored in Copi, Olive \& Schramm (1996).

Since D is converted to \he3 in the pre-main sequence phase of stellar 
evolution the chemical evolution of these two isotopes is closely linked.
In what we refer to as traditional models, \he3 may be produced or 
destroyed, though not totally (no more than what is expected in massive
stars) in the main sequence phase. As the primordial D/H ratio is 
increased, these models will produce more \he3, whose evolution must be
fit to match the solar  and present day abundances.  We discuss these abundances 
in \S 3 below.

Vangioni-Flam \etal (1994) have explored the evolution of D and \he3
using simple closed-box chemical evolution models adopting a D$_p$ of
D/H $= 7.5 \times 10^{-5}$ which is consistent with the primordial
\li7 inferred by observations and within $2\sigma$ plus 
estimated systematics of \he4.
They found that \he3 is overproduced compared with the observed solar 
and present-day values of \he3 unless it is assumed that \he3 is destroyed
significantly in low mass stars (i.e., at levels comparable to 
the destruction in more massive stars).  There exists, however, some 
observational evidence that at least some low mass stars are net 
producers of \he3. Rood, Bania, \& Wilson (1992) and Rood \etal (1995) 
find planetary nebulae with abundances of \he3 as high as $\sim 10^{-3}$.
When \he3 production in low mass stars as calculated by Iben \& Truran (1978),
Vassiladis \& Wood (1993), or Weiss, Wagenhuber, \& Denissenkov (1995),
there is always a problem with the overproduction of \he3 and in particular the
solar value of \he3/H (Olive \etal 1995; Galli \etal 1995; and 
Dearborn, Steigman, \& Tosi 1996).

As an attempt to overcome the conflict between these observations and
the results of Vangioni-Flam \etal (1994), Scully \etal (1996a)
proposed that the solution to the problem may be one of chemical
evolution.  They considered models which included a higher production of 
massive stars early in galactic history.  This was accomplished by
choosing an initial mass function (IMF) which was skewed
more towards massive star production at early times in galactic
evolution but resembled a more normal IMF at later times.
Since larger mass stars are net destroyers of \he3 this could reduce the
predicted abundance at the solar epoch. While significantly lower
solar \he3 abundances resulted from these models, they still exceeded 
the observed solar value by a factor of $\sim 2$.  They also
considered the possibility that the solar system \he3 may have been
depleted from the galactic abundance at that time and computed the 
degree to which explosions of supernovae in the solar neighborhood may
have affected the observed elemental abundances. They found that at
best only $10\%$ of the initial \he3 could be destroyed by this
process.  They concluded that chemical evolution could not fully
account for the \he3 problem.

In an attempt to find models which can reproduce the observed solar and
present-day D abundances assuming primordial deuterium values 
as high as those inferred from the
quasar absorption systems, Scully \etal (1996b) considered a class of
models which included an early epoch of star formation skewed toward 
massive star production coupled with a supernova wind-driven outflow 
mechanism.  Outflow aids in the D destruction and helps in avoiding
overproducing metals associated with the massive star production
(Vangioni-Flam \& Cass\'{e} 1995).  
These models successfully reproduce a number of observational
constraints including the age-metallicity relationship and the
distribution of low metallicity G-dwarfs.  These models, however, were 
found to overproduce \he3 by a factor of $\sim 10$ at the solar epoch
assuming standard stellar yields such as those of Iben \& Truran (1978).
The recent observations of Gloeckler \& Geiss
indicate that the sum (D + \he3)/H has remained relatively constant
since the solar epoch. This would indicate that overall low mass stars
are neither destroyers nor producers of \he3 (Turner \etal 1996).
Scully \etal (1996b) found that if stars in the narrow mass range of
.9 M$_\odot$ to $\sim 1$ M$_\odot$ produce \he3 according to standard
stellar evolutionary theory while those more massive destroy it, the
predicted \he3 abundance matches the solar and present-day
observations.  Furthermore, this would explain the high \he3
abundances measured in planetary nebulae if their progenitors lie in
this mass range.

Recent work in stellar nucleosynthesis has indicated that low mass
stars may indeed be net destroyers of \he3 if one includes
the effects of rotational mixing in low mass stars on the
red giant branch (Charbonnel 1994,
1995, Hogan 1995, Wasserburg, Boothroyd, \& Sackman, I.-J.
1995, Weiss \etal 1995, Boothroyd \& Sackman 1995, Boothroyd \&
Malaney 1995). In contrast to the result of Scully \etal 1996b, the extra  
mixing does not take place for stars which do not undergo a helium
flash (i.e. stars $>$ 1.7 - 2 M$_\odot$ ).  Thus stars {\it less than}
1.7 M$_\odot$ are responsible for the \he3 destruction.  This would
imply that the progenitor stars of the high \he3 observed by
Rood \etal must exceed this mass to avoid a conflict. Unfortunately,
the very high \he3/H abundances seen in planetary nebulae are more
representative of 1 $M_\odot$ stars, as these are the only stars able to
produce abundances as high as \he3/H $\sim 10^{-3}$. (The predicted yields
of \he3/H in low mass stars falls as $M^{-2}$ in standard stellar models
(Iben \& Truran 1978), see also Galli \etal (1996).)  

In this paper, we consider in more detail ways in which 
low mass stars affect the problem of the overproduction of \he3.
We will incorporate yields from stellar models which include 
the effects of rotational mixing (Boothroyd \& Malaney 1995) 
into a wide range of chemical evolution models to test their 
ability to solve the \he3 problem.  Note that in these models, the 
destruction of \he3 is not complete. To fit the C and O ratios, only about 80\%
destruction is required after the initial production  phase. Thus these
``destruction" models actually just keep the \he3 approximately constant which
fits the observations of Gloeckler \& Geiss (1996). As an attempt to
side-step the apparent conflict between these new yields and the
planetary nebulae abundances of \he3, we shall consider the possibility in
which some percentage of stars (on the order of 10\%)
follow a post main-sequence evolutionary path which does
not include rotational mixing.  The planetary nebulae observed by Rood
\etal would then be included in this class of stars.  We shall test
the ability of this scenario to reproduce the solar and present-day
\he3 observations.

Another possibility which we will examine is that the gas expelled
from low mass stars does not instantaneously mix in with the ISM. 
The low velocity of the planetary nebulae may translate into mixing 
times of several billions of years. Thus only a fraction of the \he3 
predicted by standard stellar evolution models for low mass stars
would be present in the gas from which the solar system formed. As we 
will see, such an assumption is not compatible with significant amounts 
of D destruction. We begin by describing our basic chemical evolution 
models.  We briefly outline the observational data of $D$ and $^3He$ 
in section 3. In section 4, we discuss in detail the results of models which
include a time delay in allowing material from planetary nebulae to
mix in with the ISM. In section 5 we discuss the results of models
which include the yields of Boothroyd \& Malaney (1995).

\section{Chemical Evolution}

In this section, we provide a basic outline for the chemical evolution
models we will be using in this work.  We want to consider models capable
of destroying deuterium by a large factor but do not overproduce
metals which generally result from the increased stellar processing
necessary.  Scully \etal (1996b) have developed a class of models which
include supernova wind-driven outflow that satisfy these criteria. In
addition, these models satisfy a number of observational constraints
such as solving the well known G-dwarf problem and reproducing the
age-metallicity relationship.  We shall adapt these models for this
work.  A summary of the main features of these models is given below.
Variations about such models such as in the stochastic models of Copi,
Schramm, \& Turner (1995b) may explain the variations in 
\he3 in Galactic \hii\ regions but do not change the core results
described here.

To follow the evolution of the mass contained in gas, we use the
classical chemical evolution equations (see e.g. Tinsley 1980).
The evolution of the gas mass density in the disk including outflow
is given by, 
\begin{equation}
\frac{dM_G}{dt}=-\psi (t)+e(t)-o(t). 
\label{gas}
\end{equation}
In this equation, $\psi (t)$ is the rate at which gas is being 
used up by star formation, $e(t)$ is the rate at which gas is returned
to the ISM by stellar deaths either in supernova events or in
planetary nebulae given by,
\begin{equation}
e(t)=\int_{m_1(t)}^{m_{up}}(m-m_R)\varphi (m,t)\psi (t-\tau (m) - T_D)dm.
\end{equation}
Here, $m_{up}$ is the upper mass limit of stars that form and
$m_R$ is the remnant mass.  We adopt the remnant mass from Iben \&
Tutukov (1984).  $\tau (m)$ is the main sequence lifetime of stars
adopted from Scalo (1986). We have also incorporated an additional
time delay, $T_D$, to account for the delay of material from planetary
nebulae in mixing in with ISM.  $T_D$ is a free parameter in our
models for stars with mass less than 8 M$_\odot$ and set equal to zero for
stars more massive than this which we assume will supernova and mix in
with the ISM on a very short time scale.

$o(t)$ in equation (\ref{gas}) is the rate at which gas leaves
the disk. In this work, we shall adopt the outflow mechanism detailed 
in Scully \etal (1996b). In theses models, some fraction of the energy of
a supernova event is assumed to go into heating ISM gas to the
escape velocity of the disk, leaving the system.  The rate at which
mass is lost from the system can be determined from,
\begin{equation}
{1\over 2}\dot{M}_W v_{esc}^2 = \epsilon E_{SN} \dot{N}_{SN},
\label{ml}
\end{equation}
where $v_{esc}$ is the escape velocity which is taken to be roughly
twice the rotational velocity i.e. $\sim$ 500 kms$^{-1}$.  We assume
that the dark matter dominates the gravitational potential of the
Galaxy and therefore the escape velocity will be taken to be a
constant. This will not be the case in merger models (see eg. Mathews
\& Schramm (1993)) where the escape velocity would be lower initially
allowing for possibly higher outflow rates.  
$E_{SN}$ is the energy per supernova event which is assumed 
to be $10^{51}$ erg and $\dot{N}_{SN}$ is the supernova rate.

All of the supernova energy would be available for heating ISM energy
to escape velocity if radiative cooling of the expanding
shell wasn't important before remnants collide.  A critical supernova
rate can be determined which the actual supernova rate must exceed if
cooling is unimportant before SNRs collide and merge.  David \etal has
determined this critical supernova rate to be,
\begin{equation}
\dot{N}_{crit}=0.83 {\rm kpc^{-3} yr}^{-1}\big({{n}\over
{\rm{cm^{-3}}}})^{1.82},
\end{equation}
where $n$ is the number density of the ISM gas.  Scully \etal (1996b)
found that in their models $\dot{N}_{SN}$ never exceeds this critical
value so cooling is always important.  Thus in order
to determine the fraction of energy per event which goes into heating 
the gas (i.e. $\epsilon$)  the residual thermal energy of an SNR after it
has collided and merged with other remnants must be determined.
Larson (1974) has estimated the ratio of the residual energy to
initial supernova energy in terms of the
supernova and critical supernova rates to be,
\begin{equation}
\epsilon = 0.22({{\dot{N}}\over {\dot{N}_{crit}}})^{0.32},
\label{ep}
\end{equation}
which is then used to determine the rate of mass loss in equation (\ref{ml}).

In addition to the gas which leaves the disk due to supernova heating,
it is assumed that some fraction of the supernovae ejecta does not
cool radiatively and is flushed directly out of the galaxy. Vader
(1986) demonstrated that simple supernova-driven wind models with 
a homogeneous ISM cannot reproduce the observed chemical properties 
of dwarf elliptical galaxies and proposed this  additional metal 
enhancement to galactic wind models. The fraction of the metals
produced in the supernova progenitors which does not cool radiatively 
and is blown out of the galaxy is denoted by $\nu$ and will be
adjusted to match the observed metallicity in the solar neighborhood.

We now turn to the evolution of heavy elements in models including
outflow of the type described above.  We assume that all stars of a mass
greater than 8 $M_\odot$ will supernova. It will be convenient to rewrite
the rate of mass ejected by stars, $e(t)$, to be the sum of the rate
ejected by those that supernova, $e_{s}$, and all other stars, $e(t)-e_{s}$.
Equation (\ref{gas}) may then be rewritten as,
\begin{equation}
\frac{dM_G}{dt}=-\psi (t) + e(t) -\nu e_s(t) - \dot{M}_W,
\label{mg}
\end{equation}
where $o(t)$ has now been replaced by $\nu e_s(t) + \dot{M}_W$ to
reflect the contribution from the wind-driven outflow and metal
enrichment. Equation (\ref{mg}) may then be extended,  
\begin{equation}
\frac{d(XM_G)}{dt}=-\psi (t)X+e_X(t) -\nu e_{sX}(t) - \dot{M}_WX,
\label{xmg}
\end{equation}
where $e_X(t)$ and $e_{sX}(t)$ represent the amount of metals 
ejected by stars and by type II supernova respectively.
This equation can be further simplified to read,
\begin{equation}
\frac{dX}{dt}=\frac{(\nu e_s(t)-e(t))X
-\nu e_{sX}(t)+e_X(t) }{M_G} 
\label{X}.
\end{equation}

As in Scully \etal (1996b)
we will consider three different initial values for 
primordial D/H.
In cases in which a high primordial D value is assumed, 
D/H = $ 2 \times 10^{-4}$ we will consider
models similar to model II in Scully \etal (1996b). The evolution of
these models is divided into two phases.  The first
phase consists of a steeply declining exponential SFR of only 
massive stars ($>$ 2 $M_\odot$) which lasts for only the first .5 - 1.0
Gyr.  This is followed by a phase with an exponential SFR with a more
normal IMF which continues to the present-day.
For the first phase, a SFR of the form $\psi (t) \sim e^{-t / \tau}$ is
chosen which lasts for $\tau$ Gyr.  A power law IMF is chosen of the
form $\varphi (m) \sim m^{-2.7}$ for the range 2 - 100 M$_\odot$.  The
second phase with a SFR $\psi (t) \sim e^{-t / 2.5}$ using the same power law
IMF but is in the range .4 - 100 M$_\odot$. 
For cases in which an intermediate D$_p$ of D/H = 7.5 $\times$
10$^{-5}$ is chosen, we shall adopt a model similar to model Ib of
Scully \etal (1996).  A SFR proportional to the gas mass is chosen,
specifically, $\psi = .28M_G$, with a power law IMF, $\varphi \propto
m^{-2.7}$ in the mass range of .4 - 100 M$_\odot$.  
Finally, we will also consider the possibility that the primordial 
D/H value is low, D/H = $ 2.5 \times 10^{-5}$. Here we will use the model
Ic from Scully \etal (1996b). In this case,  a constant SFR is assumed 
$\psi = 0.07$ with the same IMF as in the intermediate D$_p$ case.
The IMF has been
normalized in each case such that,
\begin{equation}
\int_{m_{low}}^{m_{up}}m\varphi (m)dm = 1, 
\end{equation}
where $m_{low}$ and $m_{up}$ correspond to the limits of the mass
range for each phase.

We determine the abundance of $^{16}O$ in our models adopting the
yields of Woosley \& Weaver (1995). As previously mentioned, we 
have allowed the fraction of supernova which participate in the 
galactic wind, 
denoted by $\nu$, to be a free parameter in all of our models, and
is adjusted to reproduce the solar abundance of $^{16}O$. 
In Figure 1, we show the evolution of the oxygen abundance corresponding
to the models chosen for the three values of (D/H)$_p$ and are labeled 
by this value in units of $10^{-5}$.
Also shown is the evolution of oxygen for the case of high primordial D/H
in the absence of the winds ($\nu = \epsilon = 0$) for comparison.
Since the metal enhanced wind we are employing leads to an enrichment of 
the intergalactic medium, there is a limit to the value of $\nu$ which
can be translated into a limit on the maximum value for (D/H)$_p$
in these types of models.  We will return to this question in \S 5.
All of the 
models we present unless otherwise noted result in a D evolution
consistent with the observations which we present in the next section.

\section{Observational Constraints on D and $^3He$ Evolution}

Since our primary constraint on models of galactic chemical 
evolution is the abundance of D/H (and to some extent \he3/H),
we briefly describe the adopted abundances at the present, solar,
and in the case of D/H, primordial times.
The present day ISM D abundance has been determined by Linsky \etal 
(1993, 1995) using the HST to be
\beq
{\rm (D/H)}_o = 1.60 \pm .09^{+0.05}_{-0.10}
\times 10^{-5}. 
\eeq
We shall adopt this value for D/H today.
The present day \he3 abundance has been determined in a number
of galactic \hii\ regions (Balser \etal 1994).  A large range of values
exist from $1.1 - 4.5 \times 10^{-5}$. This range may be indicative of 
a bias  and/or pollution (Olive \etal 1995) or stochastic 
evolution (Copi \etal 1995b). A recent measurement of ISM
\he3 gives (Gloeckler \& Geiss 1996)
\beq
{\rm (\he3/H)}_o = 2.1^{+.9}_{-.8} \times 10^{-5}
\eeq
We will use this latter value when comparing to the results of
chemical evolution models. The presolar D and
\he3 abundances were recently discussed in  Geiss (1992) and in
Scully \etal (1996).  Our adopted
presolar values of D,
\he3, and D + \he3 are:
\begin{equation}
[(D + ^3He)/H]_{\odot} = (4.1 \pm 0.6 \pm 1.4) \times 10^{-5},
\end{equation}
\begin{equation}
(^3He/H)_{\odot} = (1.5 \pm 0.2 \pm 0.3) \times 10^{-5},
\end{equation}
\begin{equation}
(D/H)_{\odot} = (2.6 \pm 0.6 \pm 1.4) \times 10^{-5}.
\end{equation}
We note that recent measurements of surface abundances on Jupiter
show a somewhat higher value for D/H,  D/H = $5 \pm 2 \times 10^{-5}$ 
(Niemann \etal 1996). This value
is marginally consistent with the inferred meteoritic values.

Finally, there have been several recent reported measurements of 
D/H is high redshift quasar absorption systems. Such measurements are in
principle capable of determining the primordial value for D/H and hence $\eta$
because of the strong yet monotonic dependence of D/H on $\eta$.
However, at present, detections of D/H  using quasar absorption systems
indicate both a 
high and  low value of D/H.  As such, we caution that these values may not
turn  out to represent the true primordial value. Nevertheless, we will explore
in this work, the consequences of choosing either a high or low (D/H)$_p$.
Thus we will consider in turn (D/H)$_p = 2.0 \times 10^{-4}$ (Carswell \etal
1994; Songaila \etal 1994; Rugers \& Hogan 1996a,b) 
and $(D/H)_p = 2.5 \times 10^{-5}$ (Tytler, Fan \& Burles 1996;
Burles \& Tytler 1996). We also note that the Jovian measurements of 
D/H are only marginally consistent with the low (D/H)$_p$ values.

The evolution of D in the ISM can be determined by
extending equation (\ref{xmg}). 
Since $D$ is only destroyed in stars, equation (\ref{xmg}) becomes
\begin{equation}
\frac{d(M_GD)}{dt}=-\psi (t)D - \dot{M}_W D
\end{equation}
By substitution and some further simplification, we find
\begin{equation}
\frac{dD}{dt}=\frac{(\nu e_s(t) - e(t))D}{M_G}
\end{equation}

\section{Mixing from Planetary Nebulae}

In this section, we test the possibility that \he3 is in fact produced in low
mass stars, yet the timescale for the mixing of the \he3 enriched gas into the
ISM is very slow, on the order of several Gyr.  In the limiting case, we would
withhold all of the gas ejected by stars of under $\sim $ 8
$M_\odot$ from mixing back in with the ISM.  We are motivated to
consider delaying this gas from mixing in with the ISM
since there is in fact some evidence for a dispersion in \he3 in
galactic \hii\  regions (Balser \etal 1994).  In addition, the long timescales
and relatively low matter velocity associated with planetary nebulae could
contribute to a delay in the mixing of \he3 into the ISM.

To model this effect, we first choose a primordial value for D/H.
We then run a chemical evolution model such that we delay the gas from stars
under 8 M$_\odot$ from returning to the ISM.  Figure 2 illustrates the
results for high D/H with time delays of 1, 5, and $\infty$ Gyr.  While the
\he3 evolution is now improved in each case, none is
able to reproduce a viable D and \he3 evolution, using traditional assumptions 
about \he3 evolution in low mass stars. Delay times of $<$
7.0 Gyrs can destroy enough D but still overproduce \he3.  Longer
delay times more closely reproduce the \he3 observations but do not 
destroy D by more than a factor of $\sim$ 1.5.  

Figure 3 shows similar results for model I with the lower 
(D/H)$_p$ = 7.5 $\times$
10$^{-5}$.  While a delay time of $\infty$ can reproduce the \he3
observations, again  D can not be destroyed by more than a factor of
$\sim$ 1.5. We estimate that the longest time delay which can still
reproduce both the solar and present-day D observations can only
reduce the \he3/H by a factor of 20\% from the same model run with no
time delay and traditional \he3 evolution
in low mass stars. We can conclude that this effect can clearly not be entirely
responsible for the observed solar value of \he3. This combined with our
previous results (Scully \etal 1996a and Scully \etal 1996b) all point
to the stellar yields as being primarily responsible for the observed 
\he3/H abundances. 

The exception to this conclusion is the case of low primordial D/H.
Of course in this case, D need not be destroyed by much more than a factor of 
1.5.  In Scully \etal (1996b), this case did not greatly overproduce
\he3, and a time delay of even 1 Gyr would further improve the 
comparison with the data.  It remains a difficulty in these models however,
to produce sufficient amount of heavy elements such as oxygen.

\section{Effects of New \he3 Yields}

A number of recent papers suggest that stars less massive than
2 M$_\odot$ may be net destroyers of \he3 when the effects
of rotational mixing in the red giant phase of
stellar evolution is 
 included (see e.g. Charbonnel 1994,
1995, Hogan 1995, Wasserburg, Boothroyd, \& Sackman, I.-J.
1995, Weiss \etal 1995, Boothroyd \& Sackman 1995, Boothroyd \&
Malaney 1995). Boothroyd \& Malaney (1995) give detailed results for
\he3 destruction factors which we have implemented into our chemical
evolution models. 
Main sequence mass loss is not expected to affect these yields due to the small
mass loss rates for low mass stars.
We remind the reader that these \he3 destruction mechanisms
take place in the post-main sequence phase of stellar evolution. Traditional
main sequence evolution is unchanged.
 
Boothroyd \& Malaney have already tested their results using the
models of Vangioni-Flam \etal 1994 and have adopted a model in which
$\psi = .25\sigma$.  They have suggested that values of (D + \he3)$_p$
 higher than
$1.2\times 10^{-4}$ are unable to reproduce a viable D + \he3
evolution using their \he3 yields. This corresponds to an
upper limit to (D/H)$_p < 1.0 \times 10^{-4}$. The model they have chosen, 
however, is designed to give the correct gas fraction and $D$
destruction for a (D/H)$_p = 7.5\times 10^{-5}$.  In the following, we will 
examine the evolution of D and \he3 in models which are more suited to 
destroy D for the chosen D$_p$ as well as satisfy other observational 
constraints including the present-day gas fraction using the 
Boothroyd \& Malaney yields for \he3. 

In Figure 4, we show
the results of including the Boothroyd \& Malaney yields into 
model II of Scully \etal (1996b) with high primordial D/H. 
As one can see the model has been chosen to match the solar and 
ISM D/H abundances starting at (D/H)$_p = 2 \times 10^{-4}$.
As we indicated earlier, this model incorporates supernovae driven
winds to regulate the oxygen abundance.  The evolution of \he3/H
using the Boothroyd \& Malaney yields is shown by the solid curve
in Figure 4. 
This model is able to reproduce the present day \he3/H
observation and is within the errors including systematics for the
solar observation. The effects of rotational mixing
on the red giant branch on the \he3/H abundance is dramatic as is 
seen by comparing the solid curve to the dashed one, which utilizes the
standard stellar yields of Iben \& Truran (1978).

We have also tested the new yields for the case of intermediate (D/H)$_p
=  7.5 \times 10^{-5}$.  Figure 5 shows the
resulting D/H and \he3/H evolution.  In this case, the solar \he3/H
observation is reproduced but the model gives a somewhat low
abundance of \he3/H with respect to the present day observation. Once again,
we see the significant effect of the yields by comparing the solid and 
dashed curves for \he3/H. 
For completeness, we show the results for the case of 
low (D/H)$_p = 2.5 \times 10^{-5}$ in Figure 6. In this case, because
low primordial D/H corresponds to both low $\eta$ and low primordial \he3/H,
the Boothroyd and Malaney yields predict too little \he3/H.  Indeed, the
standard yields as shown by the dashed curve show an evolution closer to the
data.  Curiously, it
seems that the reduced yields in fact work {\it better} for the higher D$_P$.   
We can conclude that the new yields based on \he3 destruction on the 
Red Giant Branch of low mass stars can solve the
\he3 problem even (and especially) for the higher D$_p$,
when combined with the Galactic evolution models of the type discussed.

Although the reduced yields of Boothroyd \& Malaney (1995)
are capable of explaining the rather flat evolution over time of \he3/H,
they can not account for the high \he3/H observed in planetary nebulae
(Rood \etal 1992,1994). As Boothroyd \& Malaney suggest, it may be that 
not all stars undergo the extra mixing subsequent to first dredge-up
and that the main-sequence produced \he3 remains intact.
In Scully \etal (1996b), it was shown that the \he3 evolution and 
planetary nebulae data could be fit if only stars between 0.9 and
$\sim 1 M_\odot$ produced \he3 in significant quantities. For the initial
mass functions considered, this represented some 10 -- 15 \%
of all stars becoming planetary nebulae. In Figure 7, we show the result
for the high D/H case when 90\% of all stars in the 1 -- 8 M$_\odot$ mass
range undergo the post main-sequence destruction while 10\% do not. It is
this 10\% which accounts for the observation of high \he3 in planetary nebulae.
Recall that all stars in this mass range produce \he3 while on the main 
sequence.  In most cases, (chosen here to be 90\%) post-main sequence
processing leads to the destruction of \he3.  In the remaining cases,
\he3 destruction might be stopped is some binary systems or in systems
in which the red giant envelope configuration is disturbed.
The dashed line in Figure 7 shows the case when 100\% of the stars destroy 
\he3 as given by the reduced yields.  As one can see, the reduced yields seem
to fit better, slightly overproducing \he3 at the solar epoch and slightly
underproducing \he3 today.  With the 10\% mix of stars which do not
destroy \he3, \he3 is overproduced at the solar epoch by a factor of about 2,
though the present day abundance is acceptable.

In Figure 8, we show the cases for (D/H)$_p = 7.5 \times 10^{-5}$
with a 90\%/10\% mix as discussed above and in Figure 9, for the case of low
D/H, the mix is 40\%/60\%.  Again, the dashed curves show the results 
using only the reduced yields.  As one can see from these figures, 
the addition of some stars which do not undergo the \he3 destruction processes 
on the red giant branch are quite consistent with the solar and
ISM observations of \he3.

Finally, as we noted earlier, a considerable amount of enriched 
material is expelled from the galaxy in these types of models, particularly
those with high primordial D/H. Such a metal enhanced wind
would contribute to the enrichment of the extra-galactic gas with
heavy elements and indeed there is evidence that such an enrichment 
has occurred.  For example, in the clusters observed by
Mushotzky \etal (1996) and  Loewenstein \& Mushotzky (1996)
the mean oxygen abundance was found to be roughly half solar.
For our model with high (D/H)$_p$, the enrichment was found to 
produce a metallicity of about $Z_{\rm IGM}\sim 0.2 Z_\odot$ 
(Scully \etal 1996b) assuming a total intergalactic gas mass equal to
ten times the Galactic mass.
 We have explored increasing (D/H)$_p$ in our models to
determine the maximum allowed value without overproducing metals in
the intergalactic medium due to our outflow mechanism.
For  (D/H)$_p = 3 \times 10^{-4}$, $Z_{\rm IGM}\sim 0.3 Z_\odot$.
In our models, we can not increase the deuterium abundance 
beyond (D/H)$_p = 5.4 \times$ 10$^{-4}$ as higher primordial abundances 
would require such an increased SFR that no gas would remain in the disk today. 
This value of (D/H)$_p$ corresponds to a $Z_{\rm IGM}\sim 0.6 Z_\odot$.

In completing this paper, we became aware of the very recent work of 
Galli \etal (1996). In that work, Galli \etal (1996) determine the
masses of the planetary nebulae observed by Rood \etal (1992, 1995).
Indeed, they confirm that the progenitor masses of the nebulae are 
$\la 2 M_\odot$, as was suspected. They also consider a chemical
evolution model with (D/H)$_p \simeq 3.5 \times 10^{-5}$ with standard stellar
yields as well as those provided by Boothroyd which account for \he3 destruction
on the red giant branch.  To match the solar and present day abundances
of \he3, they too require a mixture of these yields and their results
are in qualitative agreement with those presented here.

\section{Conclusion}   
It is clear that the evolution of \he3 is complicated by our uncertainties
in both the galactic and chemical evolution of this isotope.
It appears that the abundance of \he3 is in fact little changed over the 
history of the Galaxy. This is particularly difficult to understand
if the primordial abundance of deuterium is as high as observed in some
recent measurements of D/H in quasar absorption systems because
chemical evolution models
using standard stellar yields for \he3 in low mass stars are
known to lead to a gross overproduction of \he3 at both the solar 
and present day epochs. Unless the total deuterium
astration factor is less than about 1.5, we have
found that imposing a delay for the mixing of \he3 rich material
does not substantially improve the evolution of \he3.
This further reinforces our conclusion that chemical evolution alone
can not solve the \he3 problem.

In this paper, we have implemented new \he3 yields
for low mass stars which account for possible extra-mixing mechanisms
on the red giant branch and lead to a strong depletion of \he3
(Boothroyd \& Malaney 1996). These yields do in fact provide for a 
flat evolution for \he3 over the history of the Galaxy. However,
they can not account for the high \he3 abundances observed in 
planetary nebulae.  As such, we argue that although these extra-mixing
mechanisms may be operative for most low mass stars in their post-main
sequence evolution, in some fraction of stars the main sequence produced
\he3 must remain in tact.  In some cases, such a mix, fits the data
(solar and ISM) quite well.  We also determine an upper bound to (D/H)$_p
\la 5 \times 10^{-4}$ based on the maximum allowed
amount of extra-galactic heavy element 
enrichment.  This limit assumes that the extra-galactic
medium has less than one half solar metallicity. 
This limit is significantly higher
than the highest values of D/H observed in quasar absorption systems. In 
conclusion, the present day and solar abundances of D and \he3
can be made consistent with either high or low primordial D/H
(low or high $\eta$). Thus the solution to the the ``\he3 problem"
is clearly not cosmological but rather of Galactic or stellar origin.

\bigskip
         
{\bf Acknowledgments}  

We would like to thank M. Cass\'{e}, M. Lemoine,
and E. Vangioni-Flam for helpful discussions.  
This work  was  supported 
in part by  DOE grant DE-FG02-94ER-40823.

\newpage
\beginapjbib

\bibitem Balser, D.S., Bania, T.M., Brockway, C.J.,
Rood, R.T., \& Wilson, T.L. 1994, ApJ, 430, 667

\bibitem Boothroyd, A.I. \& Malaney, R.A. 1995, astro-ph/9512133

\bibitem Boothroyd, A.I. \& Sackman, I.-J. 1995, astro-ph/9512121

\bibitem Burles, S. \& Tytler, D. 1996, ApJ, 460,584

\bibitem Carswell, R.F., Rauch, M., Weymann, R.J., Cooke, A.J. \&
Webb, J.K. 1994, MNRAS, 268, L1

\bibitem Charbonnel,C. 1994, A \& A, 282, 811

\bibitem Charbonnel,C. 1995, ApJ, 453, L41

\bibitem Copi, C.J., Olive, K.A., \& Schramm, D.N. 1996, astro-ph/9606156

\bibitem Copi,~C.~J., Schramm,~D.~N., \& Turner,~M.~S.~1995a,  Science, 267,
192

\bibitem Copi,~C.~J., Schramm,~D.~N., \& Turner,~M.~S.~1995b,  ApJ, 455,
L95

\bibitem David, L.P., Forman, W., \& Jones, C. 1990, ApJ, 359, 29


\bibitem Dearborn, D., Steigman, G. \& Tosi, M. 1996, ApJ, 465, in press


\bibitem Epstein, R., Lattimer, J., \& Schramm, D.N. 1976, Nature, 263, 198

\bibitem Fields, B.D. \& Olive, K.A. 1996, Phys Lett B368, 103.

\bibitem Fields, B.D., Kainulainen, K., Olive, K.A., \& Thomas, D. 1996
New Astronomy, 1, 77.


\bibitem Galli, D., Palla, F. Ferrini, F., \& Penco,U. 1995, ApJ,
433, 536

\bibitem Galli, D., Stanghellini, L., Tosi, M. \& Palla 1995, astro-ph/9609184

\bibitem Geiss, J. 1993, in {\it Origin
 and Evolution of the Elements} eds. N. Prantzos,
E. Vangioni-Flam, \& M. Cass\'{e}
(Cambridge:Cambridge University Press), p. 89


\bibitem Gloeckler, G. \& Geiss, J. 1996, Nature, 381, 210

\bibitem Hogan, C.J. 1995, ApJ, 441, L17

\bibitem Iben, I. \& Truran, J.W. 1978, ApJ, 220,980

\bibitem Iben, I. \& Tutukov, A. 1984, ApJ Suppl, 54, 335

\bibitem Larson, R.B. 1974, MNRAS, 169, 229

\bibitem Lemoine, M., Schramm, D.N., Truran, J.W., \& Copi, C.J. 1996,
ApJ, in press

\bibitem Linsky, J.L., Brown, A., Gayley, K., Diplas, A., Savage, B. D.,
Ayres, T. R., Landsman, W., Shore, S. N., Heap, S. R. 1993, ApJ, 402, 694

\bibitem Linsky, J.L.,  Diplas, A., Wood, B.E.,  Brown, A.,
Ayres, T. R.,  Savage, B. D., 1995, ApJ, 451, 335

\bibitem Loewenstein, M. \& Mushotzky, R. 1996, ApJ, 466, 695

\bibitem Mathews, G.J. \& Schramm, D.N. 1993, ApJ, 404, 468

\bibitem Mushotzky, R. \etal 1996, ApJ, 466, 686

\bibitem Niemann, H.B. \etal 1996, Science, 272, 846

\bibitem Olive, K.A., Rood, R.T., Schramm, D.N., Truran, J.W.,
\& Vangioni-Flam, E. 1995, ApJ, 444, 680

\bibitem Olive, K.A., \& Scully, S.T. 1996, Int. J. Mod. Phys. A11,
409

\bibitem Olive, K.A., \& Steigman, G. 1995, ApJ S, 97, 49

\bibitem Olive, K.A., Skillman, E., \& Steigman, G. 1996, in preparation

\bibitem Pagel,~B.~J.~E., Simonson,~E.~A., Terlevich,~R.~J., \&
Edmunds,~M.~G.~1992,  MNRAS, 255, 325.

\bibitem Reeves, H., Audouze, J., Fowler, W.A., \& Schramm, D.N. 1973, 
ApJ, 179, 909

\bibitem Rood, R.T., Bania, T.M., \& Wilson, T.L. 1992, Nature, 355, 618

\bibitem Rood, R.T., Bania, T.M.,  Wilson, T.L., \& Balser, D.S. 1995, 
in {\it
 the Light Element Abundances, Proceedings of the ESO/EIPC Workshop},
ed. P. Crane, (Berlin:Springer), p. 201

\bibitem Rugers ,M. \& Hogan, C. 1996a, ApJ,  259, L1.

\bibitem Rugers ,M. \& Hogan, C. 1996b, AJ, 111, 2135

\bibitem Scalo, J. 1986, Fund. Cosm. Phys., 11, 1

\bibitem Scully, S.T. \& Olive, K.A. 1995, ApJ, 446, 272

\bibitem Scully, S.T., Cass\'{e}, M., Olive, K.A., Schramm, D.N., 
Truran, J., \& Vangioni-Flam, E. 1996a, ApJ, 462, 960

\bibitem Scully, S.T., Cass\'{e}, M., Olive, K.A., 
 \& Vangioni-Flam, E. 1996b, ApJ, in press

\bibitem Skillman, E., \etal 1996, ApJ Lett, in preparation

\bibitem Songaila, A., Cowie, L.L., Hogan, C. \& Rugers, M. 1994
Nature, 368, 599

\bibitem Spite, F. \&  Spite, M. 1982,  A.A., 115, 357

\bibitem Steigman, G., Fields, B. D., Olive, K. A., Schramm, D. N.,
\& Walker, T. P., 1993, ApJ 415, L35

\bibitem Tinsley, B.M. 1980, Fund. Cosmic Phys., 5, 287

\bibitem Turner, M.S., Truran, J.W., Schramm, D.N., \& Copi, C.J. 1996,
astro-ph/9602050

\bibitem Tytler, D., Fan, X.-M., \& Burles, S. 1996, Nature, 381, 207

\bibitem Vader, P. 1986, ApJ, 305, 669


\bibitem Vangioni-Flam, E., \& Audouze, J. 1988,
A\&A, 193, 81

\bibitem Vangioni-Flam, E. \& Cass\'{e}, M. 1995, ApJ, 441, 471

\bibitem Vangioni-Flam, E., Olive, K.A., \& Prantzos, N. 1994,
ApJ, 427, 618

\bibitem Vauclair, S. \& Charbonnel, C. 1995, A\&A, 295, 715

\bibitem Vassiladis, E. \& Wood, P.R. 1993, ApJ, 413, 641

\bibitem Walker, T. P., Steigman, G., Schramm, D. N., Olive, K. A.,
\& Kang, H. 1991 ApJ, 376, 51

\bibitem Wasserburg, G.J., Boothroyd, A.I., \& Sackmann, I.-J. 1995, ApJ,
447, L37

\bibitem Weiss, A., Wagenhuber, J., \& Denissenkov, P. 1995, astro-ph/9512120

\bibitem Woosley, S.E. \& Weaver, T.A. 1995, ApJ Supp, 101, 55

\endapjbib

\newpage   
      
\beginfig

\figitim{\bf Figure 1:} The evolution of $^{16}$O/$^{16}$O$_\odot$ for 
the chemical evolution models described in the text corresponding to
(D/H)$_p = 2 \times 10^{-4}$
with outflow (solid line) and without outflow (dotted line),
(D/H)$_p = 7.5 \times 10^{-5}$ (dashed line) and (D/H)$_p = 2.5 \times 10^{-5}$
(dot-dashed line).

\figitim {\bf Figure 2:} The evolution of D/H and \he3/H assuming a high
primordial D/H of 2 $\times 10^{-4}$.
Solid curves show the results of 
standard model \he3 yields and the chemical evolution model (II)
of Scully \etal (1996b).  Shown in this figure is the resulting evolution
when a delay is applied to the mixing time for the gas ejected by 
planetary nebulae. Delays of 1, 5, $\infty$ are shown by the dotted, dashed,
and dot-dashed curves respectively.

\figitim{\bf Figure 3:} As in Figure 2 for (D/H)$_p = 7.5 \times 10^{-5}$.

\figitim {\bf Figure 4:} The evolution of D/H and \he3/H assuming a high
primordial D/H abundance of 2 $\times 10^{-4}$.
Solid curves show the results of the
reduced \he3 yields from Boothroyd \& Malaney (1995)
and the chemical evolution model (II) of Scully \etal (1996b). The dashed 
curve shows the result using standard \he3 yields.

\figitim {\bf Figure 5:} As in Figure 4 for (D/H)$_p = 7.5 \times 10^{-5}$.

\figitim {\bf Figure 6:} As in Figure 4 for (D/H)$_p = 2.5 \times 10^{-5}$.

\figitim {\bf Figure 7:} The evolution of D/H and \he3/H assuming a high
primordial D/H of 2 $\times 10^{-4}$.
Solid curves show the results of assuming that 90\% of the stars in the 1 -- 8
M$_\odot$ range destroy \he3 as in the
reduced \he3 yields of Boothroyd \& Malaney (1995),
the remaining 10\% of stars in this mass range produce \he3 using 
standard stellar yields. The dashed 
curve shows the result from (the solid curve of) Figure 2.

\figitim {\bf Figure 8:} As in Figure 7 for (D/H)$_p = 7.5 \times 10^{-5}$.

\figitim {\bf Figure 9:} As in Figure 7 for (D/H)$_p = 2.5 \times 10^{-5}$.
Here, the mixture is 40\% reduced yields, 60\% standard yields.

\endfig

\newpage

\begin{figure}[htb]
\hspace{-1truecm}
\epsfysize=9.5truein
\epsfbox{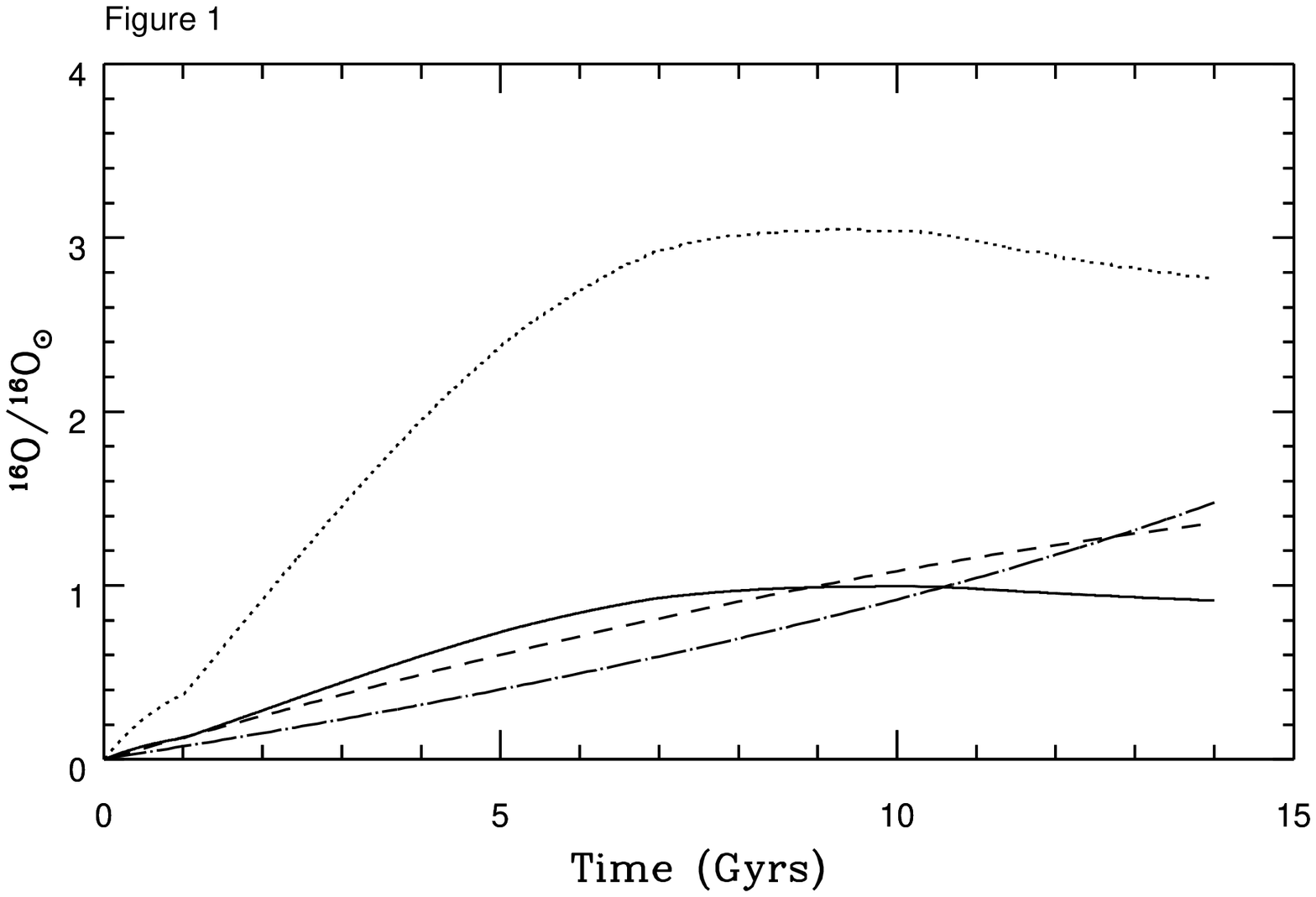}
\end{figure}  

\newpage

\begin{figure}[htb]
\hspace{-1truecm}
\epsfysize=9.5truein
\epsfbox{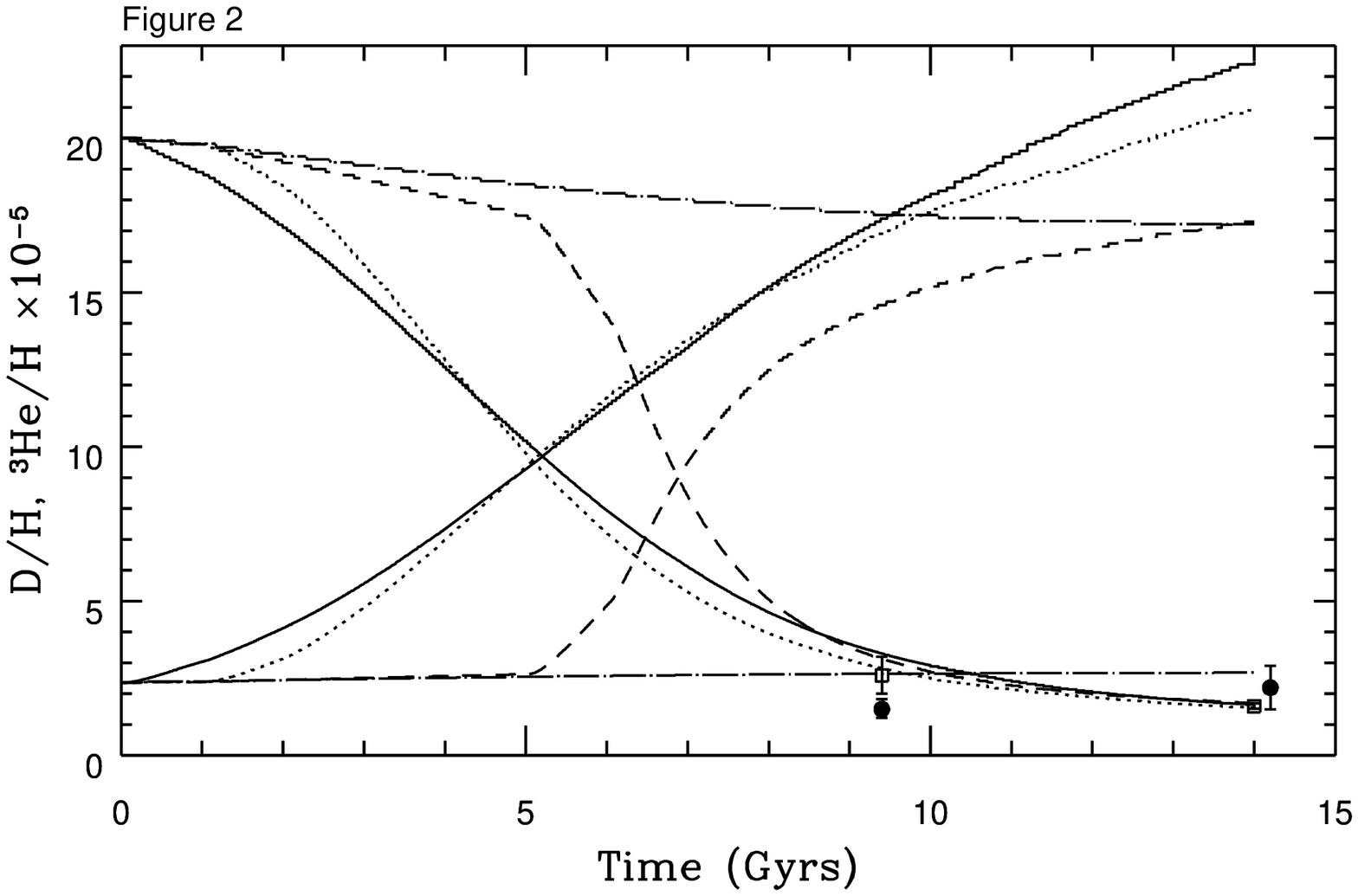}
\end{figure}

\newpage

\begin{figure}[htb]
\hspace{-1truecm}
\epsfysize=9.5truein
\epsfbox{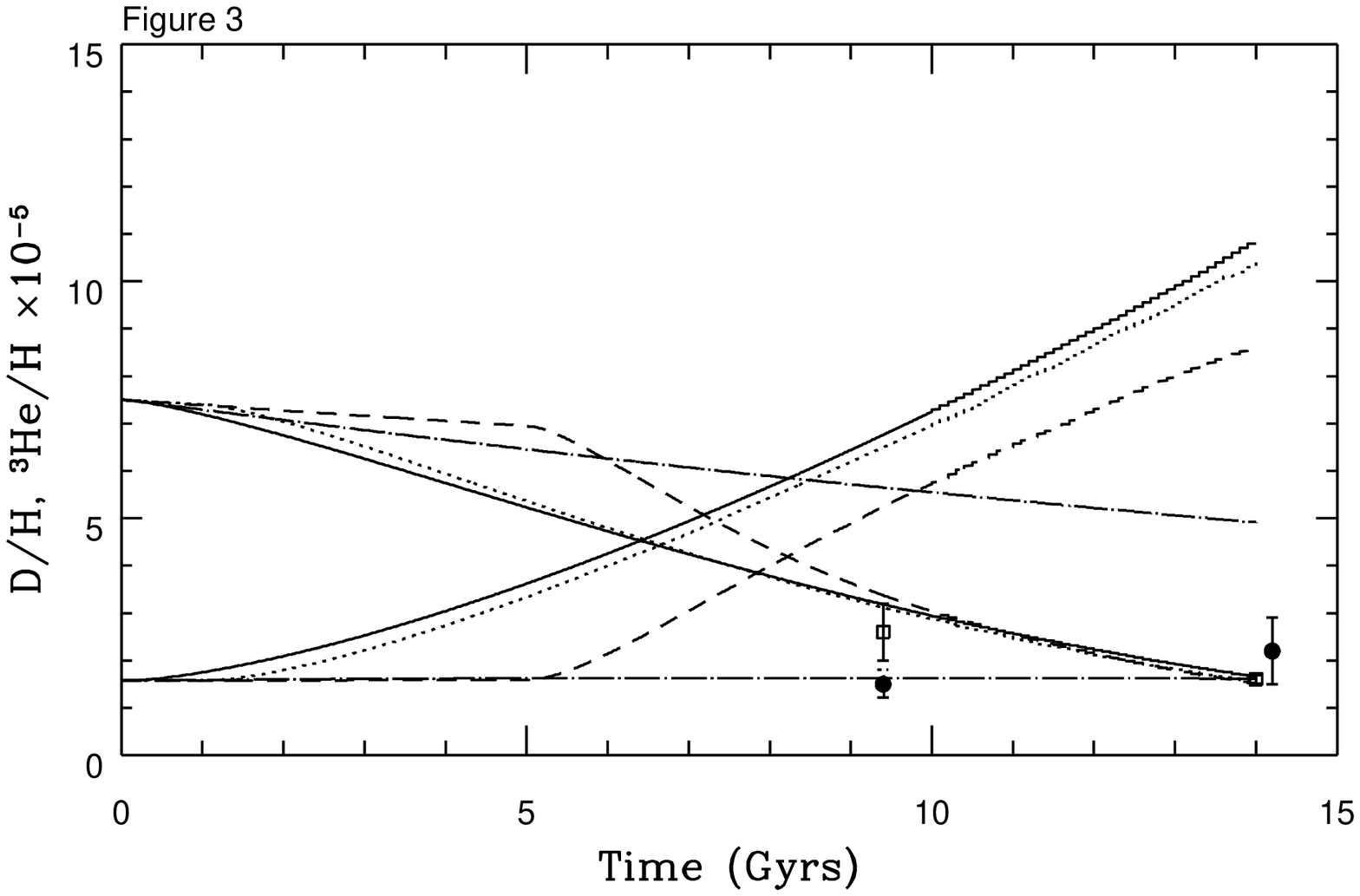}
\end{figure}

\newpage

\begin{figure}[htb]
\hspace{-1truecm}
\epsfysize=9.5truein
\epsfbox{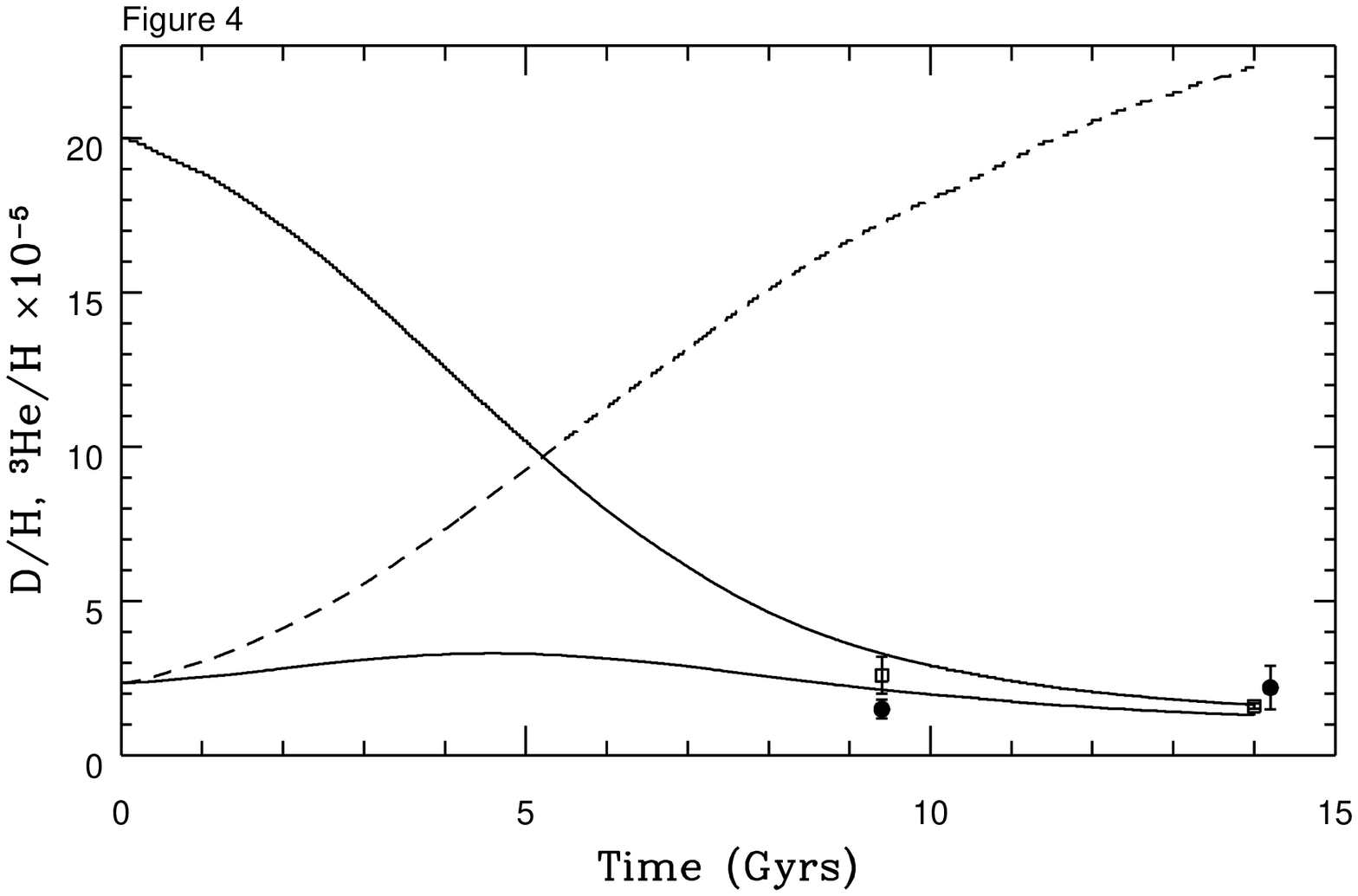}
\end{figure}

\newpage

\begin{figure}[htb]
\hspace{-1truecm}
\epsfysize=9.5truein
\epsfbox{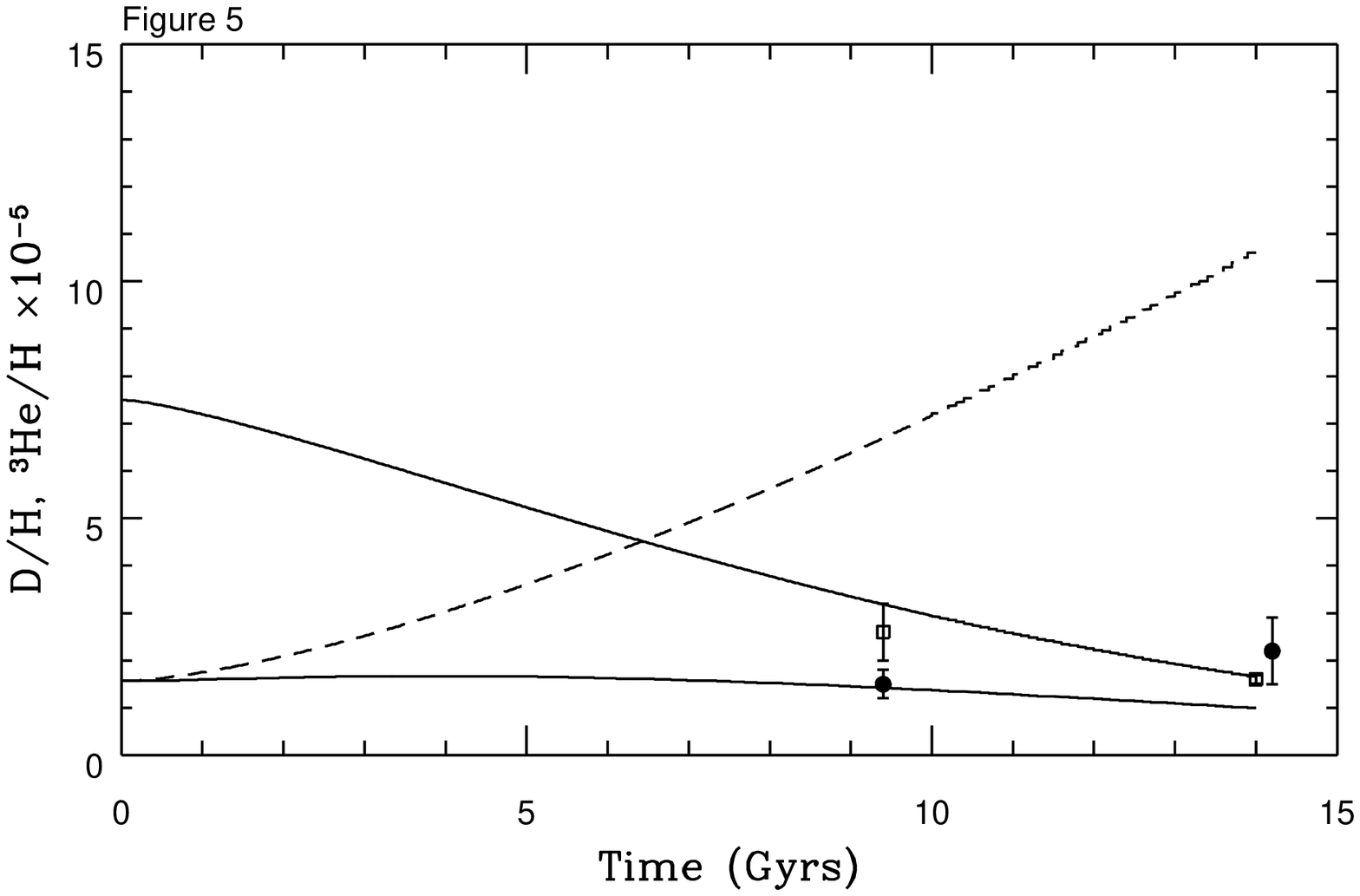}
\end{figure}

\newpage

\begin{figure}[htb]
\hspace{-1truecm}
\epsfysize=9.5truein
\epsfbox{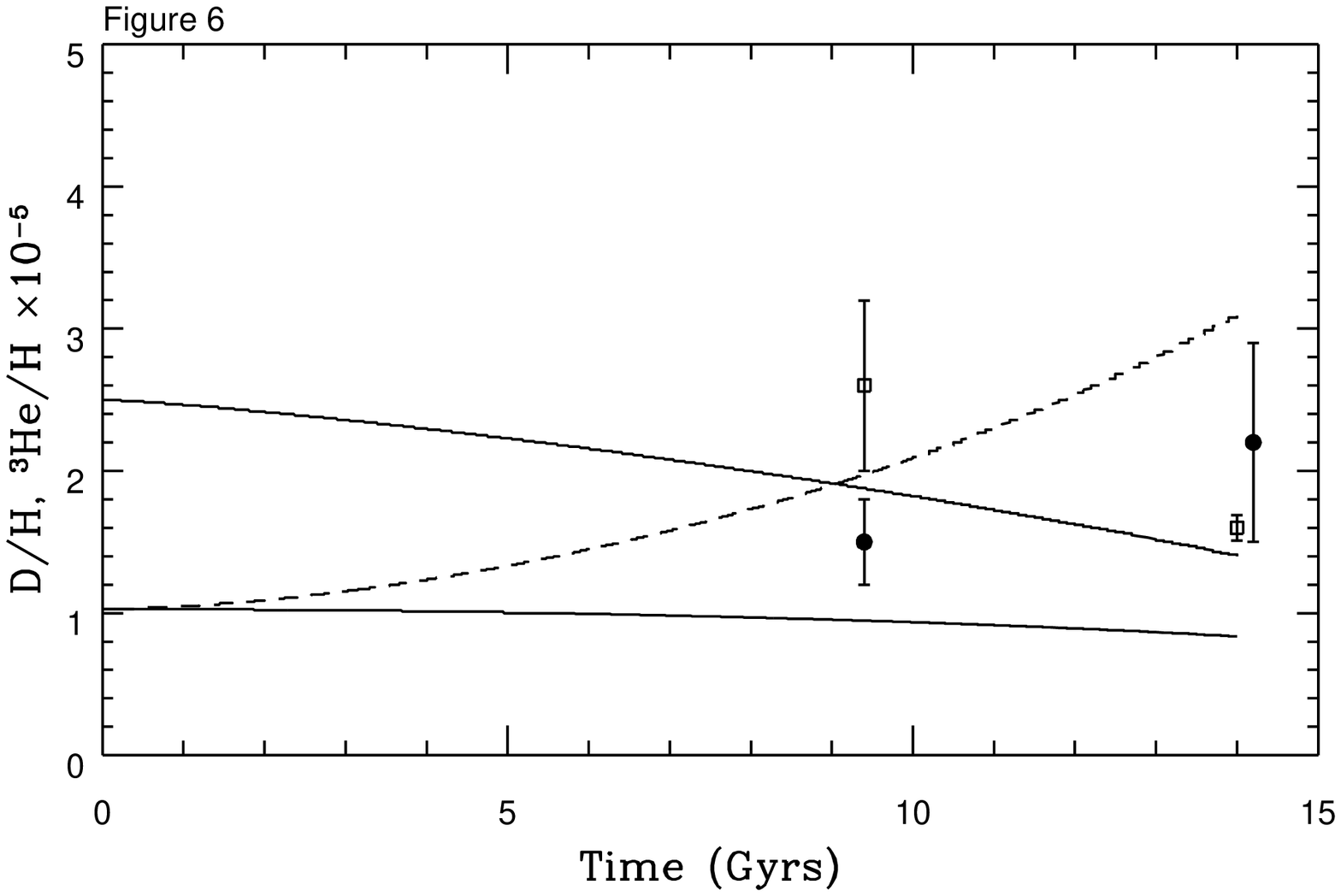}
\end{figure}

\newpage

\begin{figure}[htb]
\hspace{-1truecm}
\epsfysize=9.5truein
\epsfbox{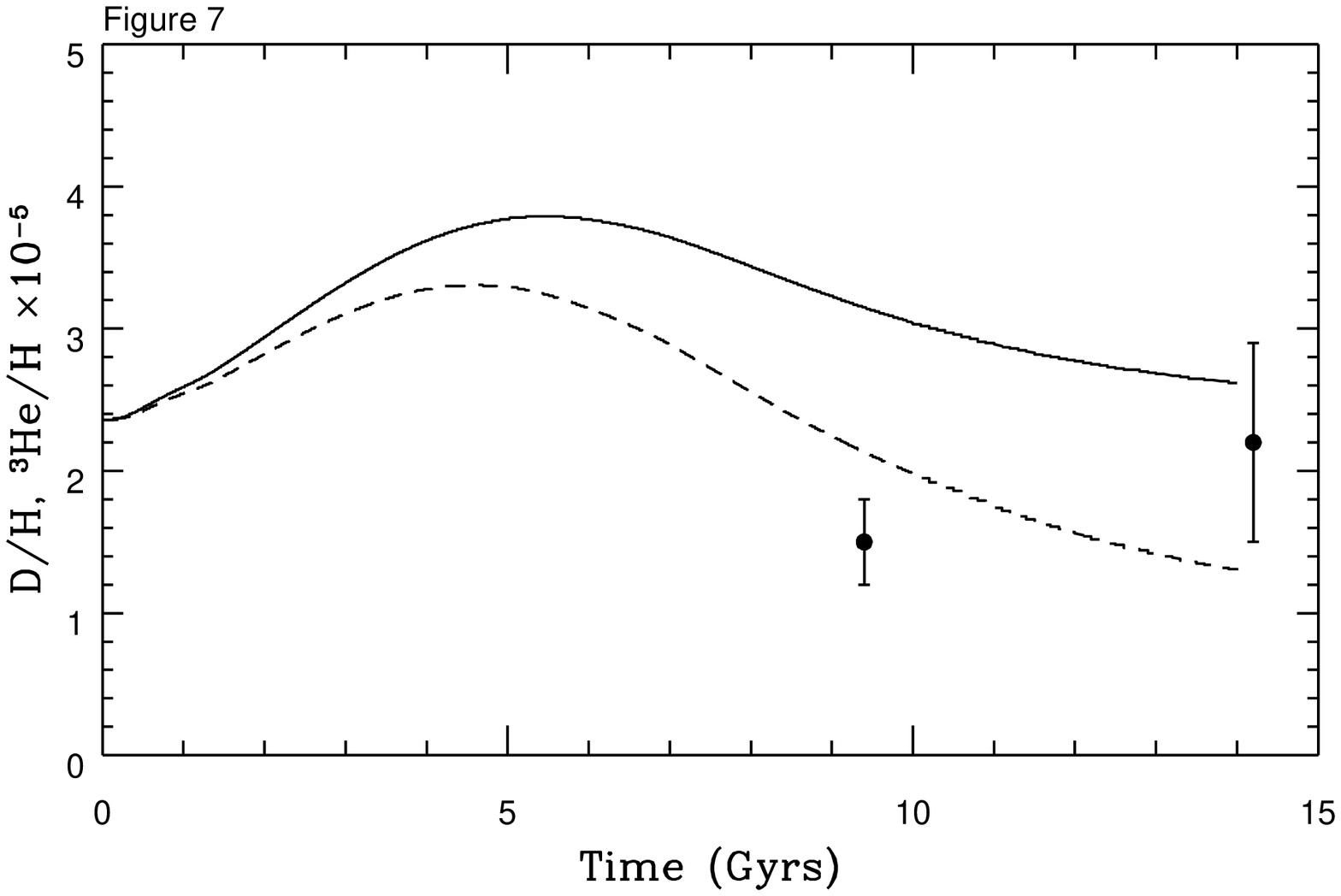}
\end{figure}

\newpage

\begin{figure}[htb]
\hspace{-1truecm}
\epsfysize=9.5truein
\epsfbox{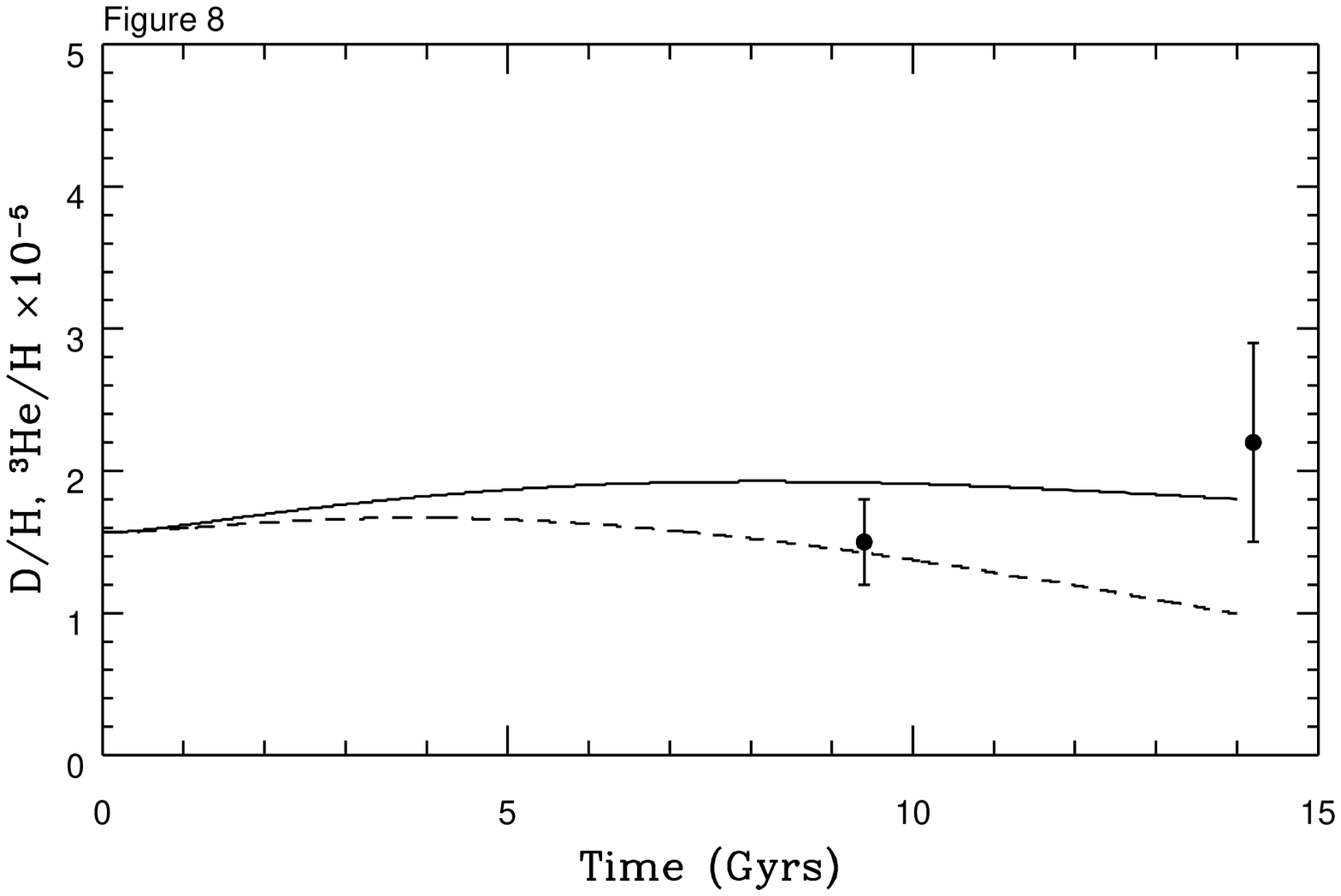}
\end{figure}

\newpage

\begin{figure}[htb]
\hspace{-1truecm}
\epsfysize=9.5truein
\epsfbox{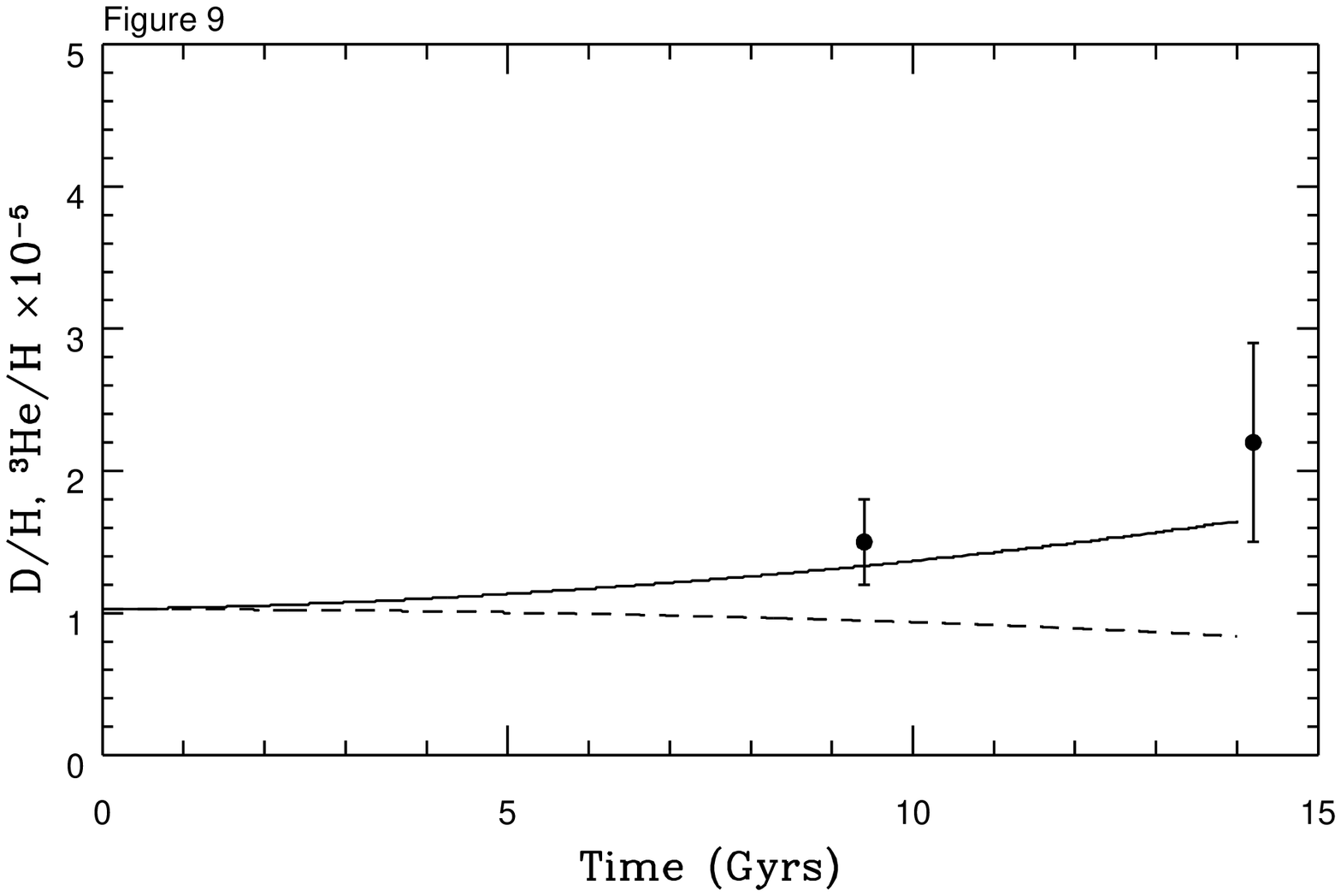}
\end{figure}

\end{document}